\documentclass[aps,prb,10pt,showpacs,preprintnumbers,amsmath,amssymb,superscriptaddress]{revtex4-2}
\usepackage{graphicx}
\usepackage{dcolumn}
\usepackage{bm}

\begin{document}

\title{Field-induced incommensurate--commensurate transitions in anisotropic magnets}

\author{P.\ T.\ Bolokhova}
\email{polina.bolokhova@gmail.com}
\affiliation{National Research University Higher School of Economics, St.\ Petersburg 190121, Russia}
\affiliation{Petersburg Nuclear Physics Institute named by B.P.\ Konstantinov of National Research Center ‘‘Kurchatov Institute’’, Gatchina 188300, Russia}

\author{A.\ V.\ Syromyatnikov}
\email{asyromyatnikov@yandex.ru}
\affiliation{Petersburg Nuclear Physics Institute named by B.P.\ Konstantinov of National Research Center ‘‘Kurchatov Institute’’, Gatchina 188300, Russia}

\date{\today}

\begin{abstract}

We study field-induced incommensurate--commensurate transitions (ICTs) in classical spin chains with Dzyaloshinskii-Moriya interaction (DMI) $D$, strong easy-plane anisotropy, and weak easy-axis anisotropy $A$ within the easy plane. This model describes systems with ferromagnetic layers coupled by exchange interaction with DMI. The magnetic field $\bf H$ is applied in the easy plane at an arbitrary angle $\delta$ to the easy axis stabilizing a soliton lattice in the ground state. We consider ferromagnetic and antiferromagnetic in-chain exchange interactions within the continuum limit and show that the easy-axis anisotropy leads to qualitatively different effects in these two cases: the ferromagnetic chain is described by a double-sine-Gordon-type equation with phase-shifted harmonics, whereas the antiferromagnetic chain reduces to a sine-Gordon equation. For the latter case, phase boundaries between incommensurate and commensurate states are found analytically in $(H,A,D,\delta)$ space and confirmed by numerical calculations for certain $D$ and $\delta$ values. Close to all ICTs, results of the continuum consideration become invalid due to a stochastic destruction of near-separatrix solutions which either make an ICT a first-order transition or stabilize states with spatially irregular soliton and antisoliton arrangements.

\end{abstract}


\maketitle

\section{Introduction}

A growing number of research is devoted to magnetic materials whose equilibrium states are spatially modulated rather than collinear. Such systems show helices, solitons, skyrmions, and other chiral or topologically nontrivial spin arrangements that can be controlled by external fields and currents, making them relevant to spintronic memory and nontraditional computing applications.~\cite{bogdanov2020,fert2017,spaldin2019,tokura2014,dong2015,vopson2015,li2021} Field-induced incommensurate--commensurate transitions (ICTs) give interesting examples of reorganization of such orderings in helimagnets whose properties are governed by the competition among exchange interactions, Dzyaloshinskii--Moriya interaction (DMI), Zeeman energy, and magnetic anisotropy. \cite{dzyaloshinskii1957,zheludevKSEA,zheludev2,BolokhovaSyromyatnikov,izyumov,iwabuchi,tensile,ross2021,boratei} In particular, it is demonstrated recently that easy-axis anisotropy can substantially reshape phase diagrams and induce transitions between modulated magnetic states.~\cite{leonovKezsmarki2017} This motivates systematic theoretical and experimental studies of field-induced transitions in the presence of a weak easy-axis anisotropy.

Starting from the work~\cite{dzyaloshinskii1957} of Dzyaloshinskii, field-induced ICTs have been well studied in a wide range of magnets with strong easy-plane anisotropy confining spins to one plane. It is standard for spin arrangements in in-plane magnetic field to be described in the continuum limit by sine-Gordon-type equations, with the structure of the incommensurate phase being of the chiral soliton lattice type. \cite{dzyaloshinskii1957,zheludevKSEA,zheludev2,BolokhovaSyromyatnikov,izyumov,iwabuchi,tensile,ross2021}
Aside from ferromagnets with weak DMI considered in Ref.~\cite{dzyaloshinskii1957}, the sine-Gordon equation appears also in antiferromagnets having weak DMI with (Ref.~\cite{zheludevKSEA}) and without (Ref.~\cite{zheludev2}) KSEA anisotropy, antiferromagnetic classical chains with strong DMI for a specific relationship between DMI and the exchange constant (Ref.~\cite{BolokhovaSyromyatnikov}), and ferromagnetic chains with easy-axis anisotropy of any order, where the magnetic field is applied along the easy axis (Ref.~\cite{izyumov}). The latter systems are described by a special double-sine-Gordon equation which can be considered analytically only in the case of the second-order anisotropy (Refs.~\cite{izyumov,iwabuchi,tensile,ross2021}).

A sine-Gordon-type equation admits uniform and soliton-lattice solutions corresponding to commensurate and incommensurate spin structures, respectively. The soliton lattice is a periodic array of solitons which are domain walls with a certain spin chirality. The ICT occurs upon field $H$ increasing when the uniform solution becomes energetically favorable compared with the soliton-lattice one.
The period of the soliton lattice of the usual sine-Gordon equation and the corresponding ground state energy behave continuously near the critical point $H=H_c$ resulting in continuous second-order phase transitions (e.g., the inverse period of the soliton lattice $\zeta \sim 1/|\ln (1-H/H_c )|$). 
It was also shown in Ref.~\cite{martynov} that higher-order derivatives in equations for the spin arrangement in antiferromagnetic systems (which were ignored in previous studies) change the type of the transition from the second-order to a first-order leading to a shift of the transition point and to jumps in physical quantities. It should be noted also that previous theoretical considerations are stimulated by corresponding experimental studies in suitable compounds. \cite{dzyaloshinskii1957,zheludevKSEA,zheludev2,BolokhovaSyromyatnikov,izyumov,iwabuchi,tensile,ross2021,boratei}

In the present study, we consider how this scenario is modified by a weak easy-axis anisotropy $A$ within the easy plane. We keep the angle $\delta$ between the easy axis and the magnetic field arbitrary, providing an extension of analysis carried out in previous considerations. The introduction of this misalignment makes even more apparent the difference between field-induced transitions in ferromagnetic and antiferromagnetic magnets. To be specific, we discuss classical chains describing systems with ferromagnetic layers coupled by weakly anisotropic ferromagnetic or antiferromagnetic exchange interaction with DMI. 

In ferromagnetic chains, a phase shift appears in the double-sine-Gordon equation at $\delta\ne0$ in which case it becomes equivalent to an equation describing the motion of a particle in a field of two traveling waves. As follows from Ref.~\cite{zaslavskii1968}, at finite $\delta$, such equation has finite instability region surrounding the separatrix in the phase space that makes the ICT a first-order transition. 

In contrast, we show for antiferromagnetic chains that the analysis carried out in Ref.~\cite{zheludev2} at $A=0$ holds up well with the replacement of $H$ with an effective magnetic field depending on $A$ and $\delta$. Then, phase boundaries between incommensurate and commensurate states can be derived explicitly from previous analytical results on the whole plane $H$--$A$ at any given $\delta$ and DMI. Accordingly, the first-order nature of ICTs in considered systems can be revealed only by incorporating higher-order derivatives in equations as it was shown for $A=0$ in Ref.~\cite{martynov}. Our numerical calculations prove the validity of all these conclusions. In particular, we recover numerically phase diagrams on the $H$--$A$ plane for some selected $\delta$ values and confirm that stochastic solutions are indeed present near phase boundaries.

The rest of this paper is organized as follows. In Sec.~\ref{theory}, we introduce the model Hamiltonian and scrutinize equations on the spin arrangement that follow from minimization of the ground state energy. The main analytical results of this paper are presented in this section, where we consider ferromagnetic and antiferromagnetic classical chains. In Sec.~\ref{numeric}, we verify numerically our analytical results and examine the near-critical behavior of antiferromagnetic chains. In Sec.~\ref{conc}, we summarize our findings.

\section{Analytical consideration}
\label{theory}

We discuss a classical spin chain described by the Hamiltonian
\begin{equation}
    \label{ham}
    \mathcal{H} = J \sum_n {\bf S}_n {\bf S}_{n+1} 
		+ A' \sum_n \left( {S}_n^y \right)^2 
		+ D \sum_n \left[{\bf S}_n \times {\bf S}_{n+1}\right]^y 
		+ A \sum_n S_n^{z'} S_{n+1}^{z'} 
		- H \sum_n S_n^z,
\end{equation}
where $0<D, A, H/S \ll |J|,A'$, strong anisotropy $A'>0$ confines spins to the $x z $ plane, the magnetic field $H$ is directed along $z$ axis, small anisotropy $A$ select the direction $z'$ in the easy plane so that the angle between $z'$ and $z$ is equal to $\delta$. We do not specify the sign of the exchange coupling constant $J$ in this section and consider the chain to be either ferromagnetic ($J<0$) or antiferromagnetic ($J>0$).

Let us denote by $\phi_n$ the angle between the spin at site $n$ and the $z$ axis, so that $S_n^z=S\cos\phi_n$. The ground state energy takes the form
\begin{equation}
    \label{ham_phi} 
    \frac{\mathcal{E}}{\tilde J} = \sum_n \left[ \pm \cos (\phi_{n+1} - \phi_n \mp \alpha) + a \cos (\phi_{n+1} - \delta) \cos(\phi_n - \delta) - h \cos \phi_n \right],
\end{equation}
where the upper and the lower signs correspond to $J>0$ and $J<0$, respectively, and
\begin{equation}
\label{const}
    \tilde J = S^2 \sqrt{J^2 + D^2}, \qquad
    \alpha = \arctan \frac{D}{|J|} \ll 1, \qquad
    a = \frac{A S^2}{\tilde J} \ll 1, \qquad
    h = \frac{H S}{\tilde J} \ll 1.
\end{equation}
Due to the smallness of $h$ and $a$, $\phi_{n+1}-\phi_{n}$ in Eq.~\eqref{ham_phi} is close to either $0$ or $\pi$ depending on the sign of $J$. This allows us to rewrite Eq.~\eqref{ham_phi} in the continuum limit, as we do below separately for ferromagnetic and antiferromagnetic exchange constants.

\subsection{Ferromagnetic chain, $J<0$}
\label{theoryFM}

The angle between spins is close to $\alpha$ at zero field and varies slowly along the chain. Then, the continuum limit of Eq.~\eqref{ham_phi} can be taken assuming that $\phi_n = \phi(n)$ is a smooth function. The extremal conditions $\partial \mathcal{E}/\partial \phi_n=0$ lead to the equation on $\phi(n)$ of the form 
\begin{equation}
    \label{phi_eq}
    \frac{d^2 \phi}{d n^2} - h \sin \phi + a \sin 2(\phi - \delta) = 0.
\end{equation}
This is similar to the double Sine-Gordon equation obtained in Refs.~\cite{izyumov,iwabuchi, tensile} except for the phase misalignment of the last term due to the non-zero angle $\delta$ between the magnetic field and the easy axis. Eq.~\eqref{phi_eq} admits uniform solutions $\cos\phi = x_i$ corresponding to commensurate phases, where $x_i$ is the $i$-th root of the equation 
\begin{equation}
	a(1-2x^2) \sin 2 \delta+ 2 a x \sqrt{1-x^2} \cos 2 \delta - h \sqrt{1-x^2} = 0.
\end{equation}
To obtain incommensurate solutions, we integrate Eq.~\eqref{phi_eq} twice with the result
\begin{equation}
    \label{phi_sol}
    n-n_0 = \mp \int_{-1}^{\cos \phi(n)} \frac{d x}{\sqrt{(1-x^2)\left(c+a (2x^2-1) \cos 2 \delta +2 a x \sqrt{1-x^2} \sin 2 \delta - 2 h x \right)}},
\end{equation}
where $c$ is a constant of integration which is to be fixed by minimization of the system energy. 

Further analysis is complicated by the term proportional to $x \sqrt{1-x^2}$ in the denominator of Eq.~\eqref{phi_sol} which is finite at $\sin2\delta \neq 0$. This term destructs analytical solutions of Ref.~\cite{tensile} obtained at $\delta=0$. Indeed, the formal substitution $n \mapsto \tilde{n} = n+ \frac{\delta}{\pi}$ in Eq.~\eqref{phi_eq} turns the term $\sin 2 (\phi(n) - \delta)$ into $\sin(2\phi(\tilde{n}) - 2 \pi \tilde{n})$ and Eq.~\eqref{phi_eq} takes the form of an equation describing the motion of a particle in a field of two waves which was investigated in Ref.~\cite{zaslavskii1968} (see Eq.~(1.2) there for $k_1/k_0 = 2$, $\Delta V_{ph} =\pi$, $\tau_0 \sim 1/\sqrt{h}$). It was shown in Ref.~\cite{zaslavskii1968} that there is a separatrix on the phase plane at $a=0$ separating solutions describing "trapped" and "free-flying" particles. At finite $a\ll h$, the last term is a perturbation in Eq.~\eqref{phi_eq} which produces a region of a stochastic instability around the separatrix the width of which depends on $\tau_0 k_1 \Delta V_{ph} \sim 1/\sqrt{h}$. In terms of system \eqref{ham}, the incommensurate phase corresponds to a "free-flying" solution of Eq.~\eqref{phi_eq} at $a=0$ whereas the commensurate phase corresponds to a solution on the separatrix. The appearance of the stochastic instability near the separatrix means that sufficiently close to $h_c$ the regular incommensurate solution can lose stability before the commensurate state is reached. The ICT can be therefore preceded near $h_c$ by a narrow stochastic regime or it can appear to be of the first order type if energies of the incommensurate and commensurate phases become equal near the stochastic instability region. \cite{bak,martynov} As follows from Ref.~\cite{zaslavskii1968}, this instability region is exponentially small in the considered case of $a \ll h$ and $h \ll 1$ ($\tau_0 k_1 \Delta V_{ph} \gg 1$). 

\subsection{Antiferromagnetic chain, $J>0$}
\label{theoryAF}

After the introduction of the slowly varying variable $\theta_n=\phi_n-\pi n$, Eq.~\eqref{ham_phi} (with upper signs) acquires the form
\begin{equation}
    \label{ham_t} 
    \frac{\mathcal{E}}{\tilde J} = \sum_n \left[ -\cos (\theta_{n+1} - \theta_n -\alpha) - a \cos (\theta_{n+1} - \delta) \cos(\theta_n - \delta) - (-1)^n h \cos \theta_n \right].
\end{equation}
Following the approach of Ref. \cite{zheludev2}, we employ the extremal conditions $\partial \mathcal{E}/\partial \theta_n = 0$ to obtain equations for $\theta_n$. In process, we take advantage of the fact that $\theta_{n+1} - \theta_n - \alpha \ll 1$ since $a,h \ll 1$. The extremal conditions give finite-difference equations
\begin{equation}
    \label{theta_dif}
    \theta_{n+1}-2\theta_n +\theta_{n-1} - a \sin 2 (\theta_n - \delta) - (-1)^n h \sin \theta_n = 0.
\end{equation}
We search for a solution in the form $\theta_n = \Theta (n) + (-1)^n \vartheta (n)$, where $\Theta(n)$ and $\vartheta (n)$ are smooth functions. Then, one obtains from Eq.~\eqref{theta_dif} in the leading order in $h$ and $a$ within the continuum approximation
\begin{subequations}
\label{theta_eqs}
\begin{eqnarray}
\label{theta_eqs1}
    \frac{d^2 \Theta}{d n^2} &=& - \frac{h_{\mathrm{eff}}^2}{8} \sin 2(\Theta+\chi),\\
\label{theta_eqs2}
    \vartheta &=& - \frac{h}{4} \sin \Theta,
\end{eqnarray}
\end{subequations}
where
\begin{equation}
    \label{h_eff}
    h_{\mathrm{eff}} = \left(h^4 - 16 a h^2 \cos 2 \delta + 64 a^2\right)^{1/4},
\end{equation}
\begin{equation}
    \label{chi}
    \chi = \frac{1}{2} \operatorname{arccot} \frac{h^2 - 8 a \cos 2 \delta }{8 a \sin 2 \delta},
\end{equation}
and we take the codomain of $\operatorname{arccot}$ to be $\left[0; \pi\right)$, so that $\chi \in \left[0; \pi/2\right)$. A special case is $\delta = 0$, when $\chi = 0$ for $h^2 \geq 8a$ and $\chi = \pi/2$ for $h^2 < 8a$. It is seen from Eq.~\eqref{h_eff} that $h_{\mathrm{eff}}$ is finite at $h=0$ that signifies the formation of the soliton lattice by the easy-axis anisotropy $a$ at zero field.

Note that in contrast to the double-sine-Gordon equation \eqref{phi_eq} appearing in ferromagnetic systems with easy-axis anisotropy \cite{iwabuchi,tensile}, Eq.~\eqref{theta_eqs1} is a regular sine-Gordon equation with the shifted double argument which is similar to the one obtained in Ref.~\cite{zheludev2} for antiferromagnetic systems without easy-axis anisotropy. The origin of this difference is that the staggered substitution maps both the easy-axis anisotropy and the field-induced contribution onto second harmonics of the slowly varying antiferromagnetic phase. As a result, their leading-order contributions combine into a single shifted sine term in Eq.~\eqref{theta_eqs1}. 

\begin{figure}[t]
    \centering
    \includegraphics[width=0.3\linewidth]{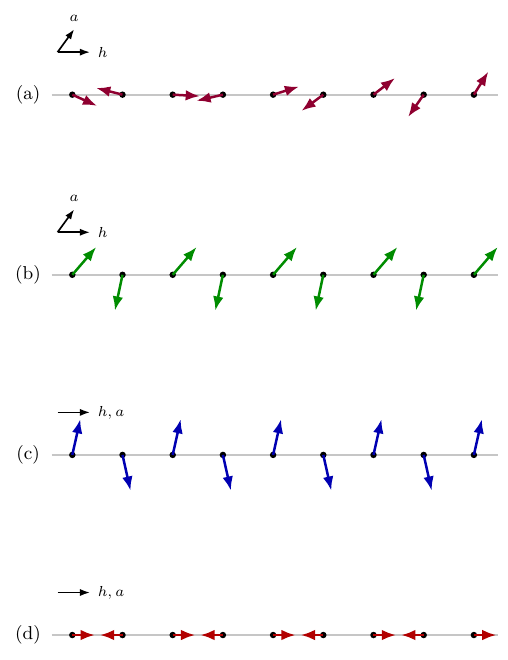}
    \caption{Schematic spin configurations in the $xz$ plane of antiferromagnetic chain \eqref{ham}. (a) Typical incommensurate arrangement. (b) Typical commensurate (spin-flop) arrangement for a misaligned easy axis $a$ and the field $h$ (the angle between the field and the easy axis $\delta = 3\pi/10$ for definiteness). (c) Commensurate (spin-flop) configuration at $\delta=0$ and $h^2 > 8a$. (d) Commensurate (antiferromagnetic) configuration at $\delta=0$ and $h^2 < 8a$.}
    \label{fig:spin-configurations}
\end{figure}

Further analysis is similar to that in Ref.~\cite{zheludev2}, except that we have to take into account the shift by $\chi$ in Eq.~\eqref{theta_eqs1}. Eqs.~\eqref{theta_eqs} admit a non-soliton solution $\Theta = \pi/2 - \chi$, $\vartheta = -\frac{h}{4} \cos \chi$ which corresponds to the spin-flop phase except for the case of $\delta=0$, $h^2<8a$, when $\chi=\pi/2$ and the non-soliton solution describes the antiferromagnetic arrangement. The general solution of Eqs.~\eqref{theta_eqs} has the form
\begin{eqnarray}
    \label{theta_sol}
    \Theta(n) &=& -\chi + \text{am} \left( n \frac{h_{\mathrm{eff}}}{2 \beta}, \beta \right), \\
    \vartheta(n) &=& - \frac{h}{4} \left[ \cos \chi\ \text{sn} \left( n \frac{h_{\mathrm{eff}}}{2 \beta}, \beta \right) - \sin \chi\ \text{cn} \left( n \frac{h_{\mathrm{eff}}}{2 \beta}, \beta \right) \right],\nonumber
\end{eqnarray}
where $\text{am} (x, \beta)$, $\text{sn} (x, \beta)$ and $\text{cn}(x, \beta)$ are Jacobi elliptic functions of modulus $\beta$. Spin configurations in all solutions mentioned above are shown schematically in Fig.~\ref{fig:spin-configurations}. 

The system energy per spin derived in the continuum limit from Eq.~\eqref{ham_t} with the use of Eqs.~\eqref{theta_eqs} has the form 
\begin{equation}
    \label{ham_Theta} 
    \frac{\mathcal{E}_0}{\tilde J} = \frac{(\Theta'-\alpha)^2}{2} + \frac12 \frac{h_{\mathrm{eff}}^2}{8} \sin 2 (\Theta + \chi),
\end{equation}
which is equivalent to the expression for the energy in the absence of the easy-axis anisotropy with the substitution $\Theta \mapsto \Theta+\chi$ and $h \mapsto h_{\mathrm{eff}}$. 
Therefore, we can rely on the analysis of the ICT carried out in Refs.~\cite{zheludev2,martynov}. In the continuum approximation, the continuous ICT occurs at 
\begin{eqnarray}
\label{hc}
    h^{(cr)}_{\mathrm{eff}} = \pi \alpha
\end{eqnarray}
with the incommensurability parameter $\zeta(H,A)$ decreasing to zero as $1/|\ln (h_{\mathrm{eff}}^{(cr)}-h_{\mathrm{eff}})|$ ($\zeta(H,A)$ is the inverse period of the soliton lattice). However it was shown in Ref.~\cite{martynov} that the ICT is of the weak first-order type outside the continuum approximation (the type of the ICT changes when one takes into account higher-order derivatives in Eqs.~\eqref{theta_eqs}). It was demonstrated in Ref.~\cite{martynov} that a region of chaotic solutions arises near $h^{(cr)}_{\mathrm{eff}}$ and the commensurate phase becomes energetically more favorable than the incommensurate one before reaching this region. We demonstrate in the next section the modification of the ICT by numerical solution of lattice equations.

\section{Numerical results for antiferromagnetic chain}
\label{numeric}

We compare in this section analytical results of Sec.~\ref{theoryAF} obtained in the continuum approximation with our numerical findings for the antiferromagnetic chain. In order to obtain accurate numerical solutions, we numerically solve the following exact set of equations derived from minimization of energy \eqref{ham_t}:
\begin{equation}
    \label{theta_dif_sin}
    \sin(\theta_{n+1}-\theta_n-\alpha) -\sin(\theta_n - \theta_{n-1}-\alpha) - a \sin 2 (\theta_n - \delta) - (-1)^n h \sin \theta_n = 0.
\end{equation}
First, we verify the analytical transition condition numerically for several representative values of $\delta$ taking $\alpha=0.01$ (see  Eq.~\eqref{const}). For this, we consider a chain with $N\sim10^5$ spins with open boundary conditions. It is seen that once the values of $\theta_1$ and $\theta_2$ are specified, one can easily determine the remaining $\theta_n$ from Eqs.~\eqref{theta_dif_sin}. Then, the incommensurate solutions of Eqs.~\eqref{theta_dif_sin} can be parametrized by $\theta_1$ and $\theta_2$. For each set of parameters $(h,a,\delta)$, we substitute the solution of Eqs.~\eqref{theta_dif_sin} obtained in this way into Eq.~\eqref{ham_t} for the energy, then minimize the energy with respect to $\theta_1$ and $\theta_2$ and compare the resulting ground state energy with the energy of the commensurate spin-flop phase $\theta_n = \pi/2 - \chi-(-1)^n \frac{h}{4} \cos \chi$, where $\chi$ is given by Eq.~\eqref{chi}. If the commensurate state has lower energy, we identify it as the ground state for these parameters. 

\begin{figure}[t]
    \centering
    \begin{minipage}{0.48\linewidth}
        \centering
        \includegraphics[width=\linewidth]{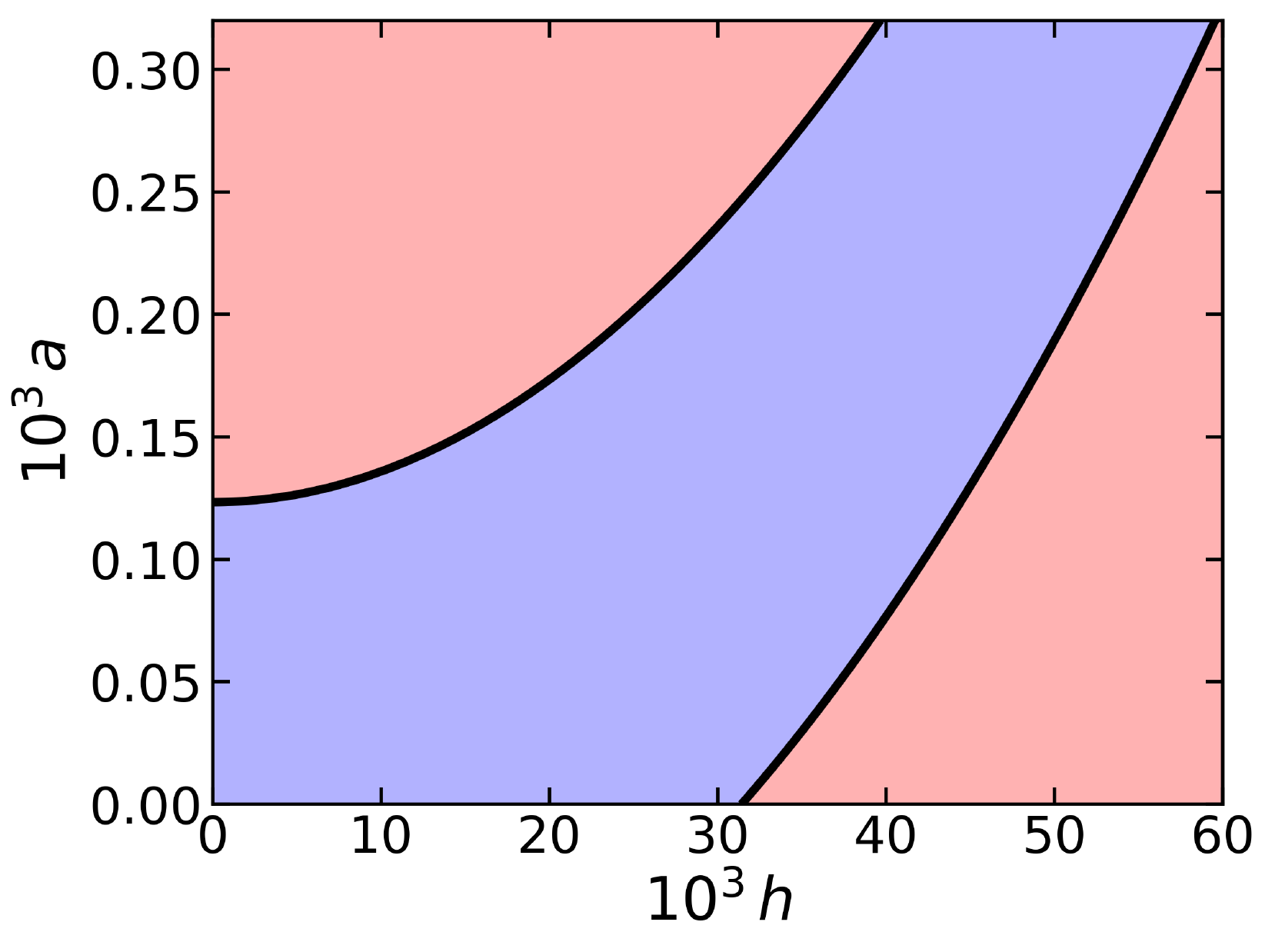}
        \vspace{0.5ex}
        (a) $\delta=0$
    \end{minipage}\hfill
    \begin{minipage}{0.48\linewidth}
        \centering
        \includegraphics[width=\linewidth]{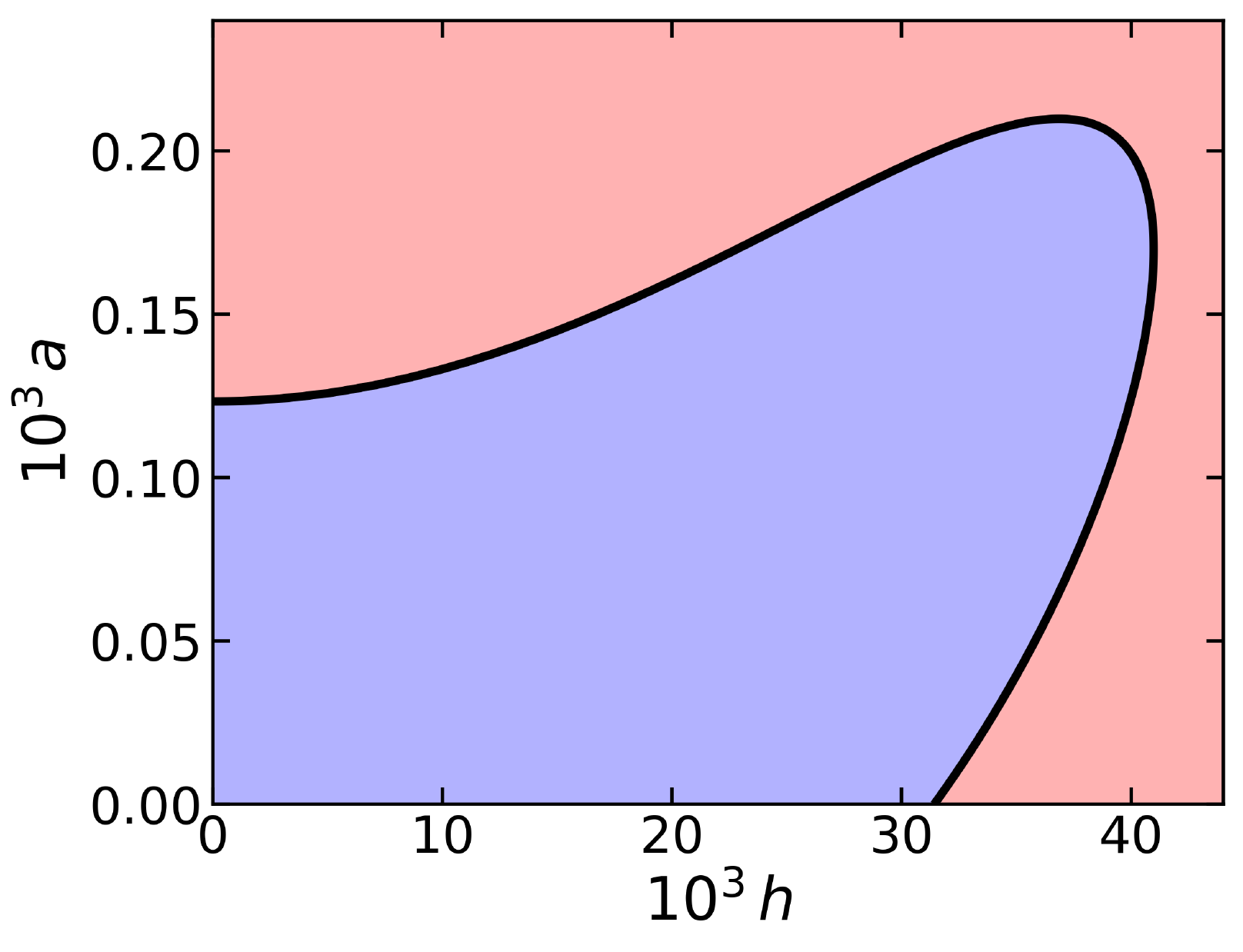}
        \vspace{0.5ex}
        (b) $\delta=\pi/10$
    \end{minipage}
    \vspace{1ex}

    \begin{minipage}{0.48\linewidth}
        \centering
        \includegraphics[width=\linewidth]{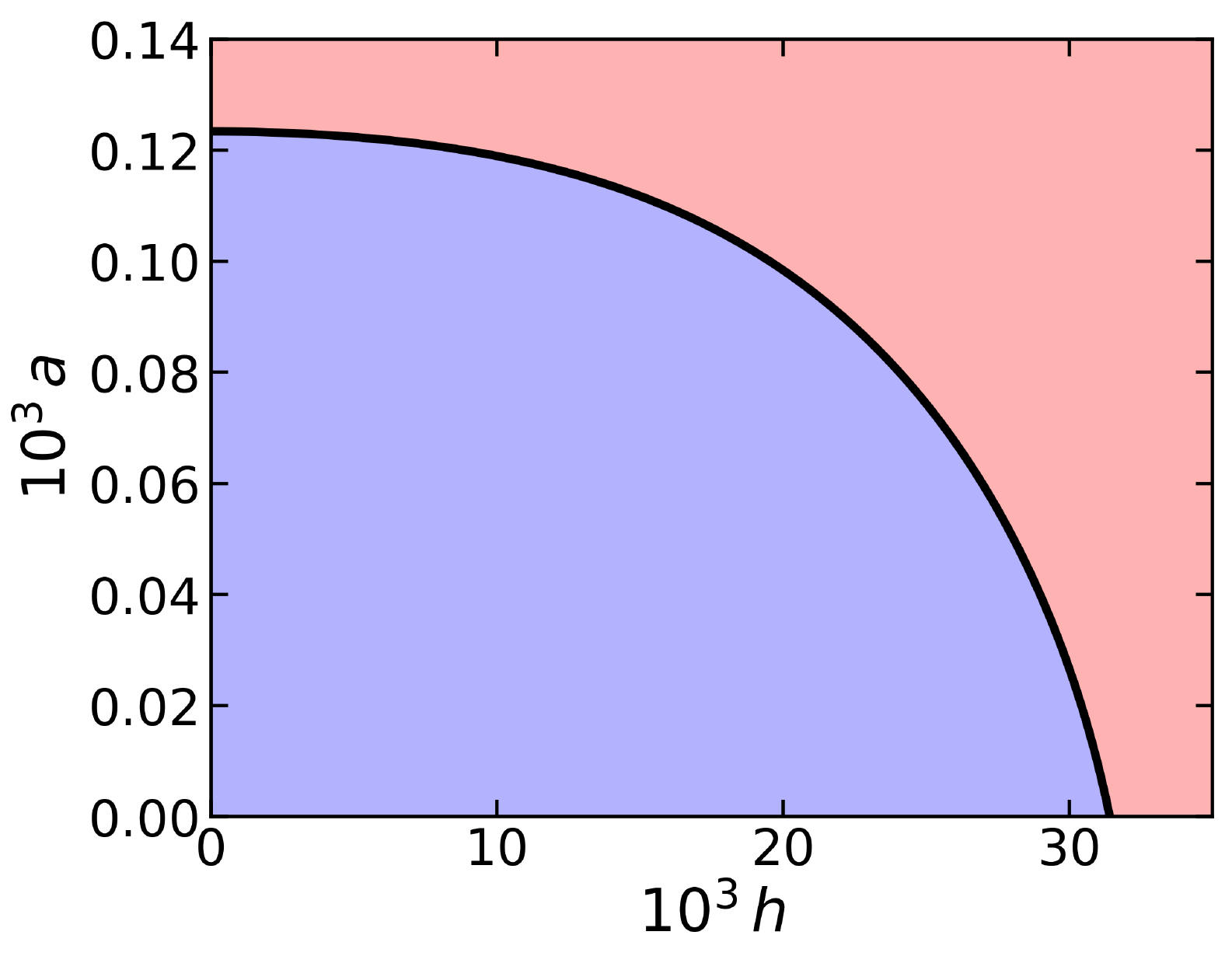}
        \vspace{0.5ex}
        (c) $\delta=3\pi/10$
    \end{minipage}\hfill
    \begin{minipage}{0.48\linewidth}
        \centering
        \includegraphics[width=\linewidth]{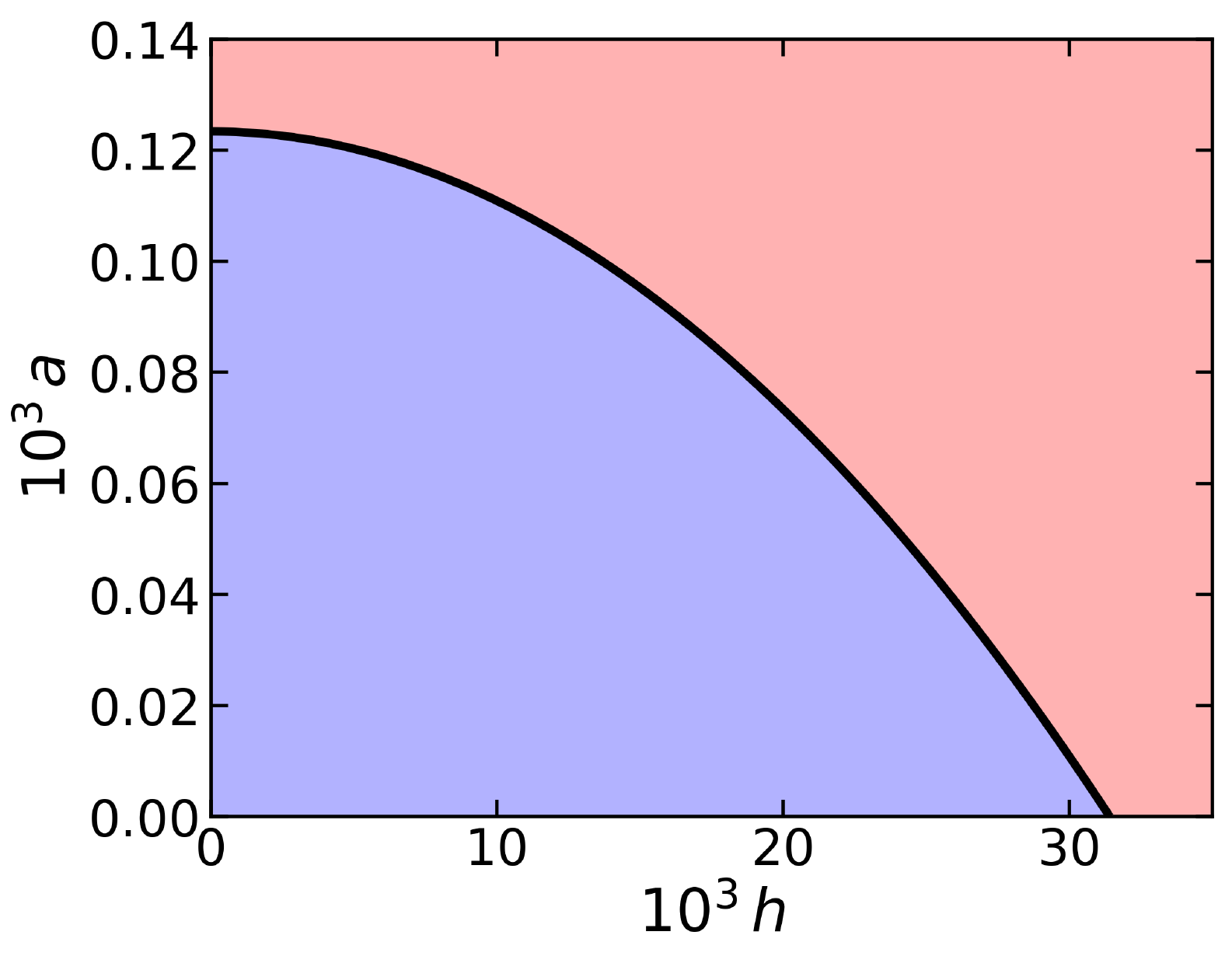}
        \vspace{0.5ex}
        (d) $\delta=5\pi/10$
    \end{minipage}
    \caption{Phase diagrams of the antiferromagnetic chain obtained analytically in the continuum approximation from Eq.~\eqref{hc} and confirmed numerically for different angles $\delta$ between the magnetic field $\bf h$ and the easy axis. Here $\alpha=0.01$ defines the DMI value (see Eq.~\eqref{const}). Pink and violet regions correspond to commensurate and incommensurate phases, respectively. Typical spin arrangements in these phases are shown in Fig.~\ref{fig:spin-configurations}.}
    \label{fig:phase-diagrams}
\end{figure}

Fig.~\ref{fig:phase-diagrams} shows phase diagrams on the $h$--$a$ plane which we find from Eq.~\eqref{hc}, where $h_{\mathrm{eff}}$ is given by Eq.~\eqref{h_eff}. The resulting numerical transition fields for the chain length $N=5 \cdot 10^5$ deviate from analytical prediction by less than $0.1\%$ (this difference is visually indistinguishable on the diagrams). Increasing the chain length, as well as expanding the set of initial conditions $(\theta_1, \theta_2)$ parametrizing the solutions, further reduces the discrepancies. This is due to the fact that solutions become sensitive to initial conditions near the critical point, and increasing the system size helps to suppress the effect of this on numerically obtained characteristics. However, a qualitative difference between numerical results and analytical transition does emerge close to the transition point as we show below by considering very long chains.

\begin{figure}[t]
\centering
\includegraphics[width=0.6\linewidth]{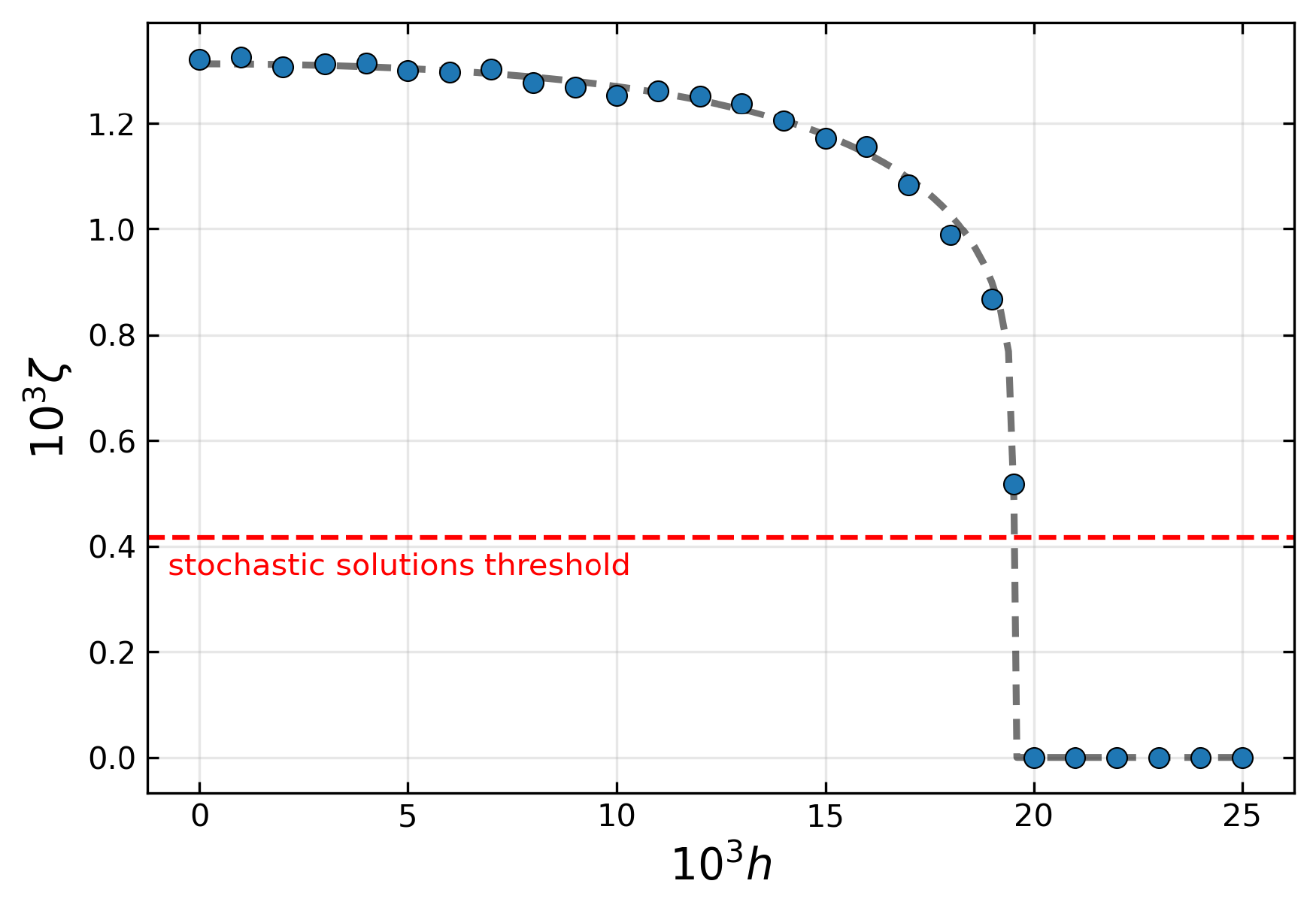}
\caption{
Incommensurability parameter $\zeta$ (the inverse period of the soliton lattice) as a function of the magnetic field $h$. Points show numerical findings obtained from finite-difference solutions and the dashed line represents the analytical solution in the continuum approximation. Very close to the ICT, the incommensurate phase becomes unstable against a spatially irregular soliton state so that $\zeta$ cannot be smaller than a critical value shown by the horizontal dashed line (see the text).}
    \label{fig:zeta-h}
\end{figure}

Next, we determine the dependence of the magnetic propagation vector $\zeta(h,a)$ of the incommensurate state on the system parameters. It is defined as the inverse period of the soliton lattice and, as predicted by continuum theory, is expected to decrease to zero logarithmically as the system approaches the critical point and solitons move apart from each other. Here, we consider a chain with $N=5 \cdot 10^5$ spins, while $\alpha$ is fixed at $0.01$. As follows from Eq.~\eqref{ham_Theta}, the state of the system depends on the parameters through $h_{\mathrm{eff}} (h,a,\delta)$, and here we investigate $\zeta(h_{\mathrm{eff}})$ by varying $h$ at fixed $a=0.0001$ and $\delta=3\pi/10$. We use the algorithm described above to find ground states for different values of $h$, and then calculate the inverse soliton period of states belonging to the incommensurate phase. The resulting dependence $\zeta(h)$ is plotted in Fig.~\ref{fig:zeta-h}. The dashed grey line shows the analytical prediction obtained from the inverse period of solutions \eqref{theta_sol}. The agreement with the theoretical curve is also good, and no jump in $\zeta(h)$ is resolved for parameter values used in the calculations. However, a closer look at solutions near the transition point by increasing the chain length up to $N=10^7$ reveals spatially irregular solutions at $1-h_{\mathrm{eff}}/h_{\mathrm{eff}}^{(cr)} \sim 10^{-5}$ whose energies are lower than those of regular soliton-lattice solutions and which contain chaotic arrangement of solitons and antisolitons (i.e., solitons with opposite spin chirality). This suggests that the continuum approach ceases to describe the ground state just near the critical point.

To describe the physical nature of this disruption as well as other features near the transition point, we refer to a very similar Frenkel--Kontorova model, chaotic critical behavior of which is described in Refs.~\cite{bak,braunKivshar2004}. In a discrete chain, the energy of a soliton can depend on its position relative to the lattice sites producing a Peierls--Nabarro pinning barrier. As the soliton separation increases near the transition, their mutual interaction weakens and even weak lattice pinning may become important for their arrangement. For this reason, the continuum limit ceases to describe solutions accurately. The regular incommensurate solution becomes unstable and the system either enters a stochastic regime or switches directly to the commensurate state. In other words, the discreteness of the chain modifies the system's behavior close to the transition, giving it features of the first-order transition, but it does not affect the validity of continuum solutions outside this narrow region. \cite{bak, martynov, braunKivshar2004}

The value of $\zeta$ of the incommensurate solution immediately before it becomes unstable is indicated by the red dashed line in Fig.~\ref{fig:zeta-h}. Below this level, the analytical curve represents just a continuation of the continuum solution. It would be practically impossible to distinguish the threshold value of $h_{\mathrm{eff}}$ experimentally due to its extreme closeness to $h_{\mathrm{eff}}^{(cr)}$ ($1-h_{\mathrm{eff}}/h_{\mathrm{eff}}^{(cr)} \sim 10^{-4}-10^{-6}$). Notice that a similar estimate for the threshold value of $h_{\mathrm{eff}}$ was obtained previously in Ref.~\cite{martynov} in a system without easy-axis anisotropy.

\section{Conclusions}
\label{conc}

To conclude, we analyze spin arrangements and field-induced incommensurate--commensurate transitions (ICTs) in one-dimensional ferromagnetic and antiferromagnetic classical chains with the Dzyaloshinskii--Moriya interaction and the weak easy-axis anisotropy in the easy plane. Keeping the angle $\delta$ between the magnetic field and the easy axis arbitrary, we show that the ferromagnetic and antiferromagnetic cases lead to qualitatively different equations on the spin arrangement in the continuum approximation. In the ferromagnetic case, the field and anisotropy generate a double-sine-Gordon-type equation with phase-shifted harmonics (see Eq.~\eqref{phi_eq}). In the antiferromagnetic case, the leading equation reduces to a shifted sine-Gordon equation with a double argument (see Eq.~\eqref{theta_eqs1}) in which an effective magnetic field $h_{\mathrm{eff}}$ Eq.~\eqref{h_eff} arises. This allows to derive phase boundaries between commensurate and incommensurate states in the form Eq.~\eqref{hc}. This result extends the standard continuum description of the field-induced transition by incorporating an arbitrary in-plane easy-axis anisotropy.

Numerical solutions of the finite-difference equations for long antiferromagnetic chains and for selected values of $\delta$ confirm analytical phase diagrams shown in Fig.~\ref{fig:phase-diagrams}. However, numerical solutions reveal an instability of the incommensurate state very close to phase boundaries against irregular soliton solutions. This effect may be attributed to the stochastic destruction of near-separatrix solutions and discrete pinning effects such as a Peierls--Nabarro barrier.

Thus, results of this work confirm that easy-axis anisotropy in helical magnets modifies field--induced transitions in ways that depend on the intrinsic symmetries of the system. In ferromagnetic chains, the anisotropy competes directly with the magnetic field and enhances the mechanisms leading to stochastic destruction of incommensurate solutions near ICTs. In antiferromagnetic chains, by contrast, the leading effect of the easy-axis anisotropy is absorbed into a renormalization of the magnetic field value, so that the standard sine-Gordon description remains applicable over most of the phase diagram. This distinction serves as an illustration for how small low-symmetry spin interactions such as anisotropy may affect ICTs in different classes of helical magnets.

\bibliography{references}

\end{document}